
\input phyzzx


\nopubblock
\PHYSREV
\parindent=0 truecm
\parskip=0 truecm
\titlepage
{\bf
\title{TESTS OF GENERAL RELATIVITY AT LARGE DISTANCES AND DARK
MATTER\foot{Invited Talk to be published in the Proceedings of Rencontre
Du Vietnam on Particle Physics And Astrophysics, Hanoi,
December 1993}}
\author{Arnon Dar\foot{ Spported in part by the Technion
 Fund for Promotion of Research}}
\address{Department of Physics and Asher Space Research Institute
Israel Institute of Technology, Haifa 32200, Israel}
\abstract
\baselineskip=8pt
The most simple observed cases of gravitational lensing of
distant quasars and galaxies by galaxies and clusters of galaxies are
used to test Einstein's theory of General Relativity
and Newtonian Gravity over galactic and intergalactic
distances. They extend the
distance range over which Newton-Einstein Gravity
has been tested by 10 orders of magnitude.
Although the precision of the tests are far from the precision
of the solar system tests of EGR and those from pulsar timing
in close binary systems, they confirm quite accurately
the validity of Einstein's General Relativity and its weak
field limit, Newtonian Gravity, over galactic and intergalactic
distances and the  existence of large quantities of dark matter
in galaxies and clusters of galaxies.
Future observations can improve
the accuracy of the tests and reduce the possibilities for
systematic errors.}
\endpage
\parskip=0 truecm
\baselineskip=8pt

{\bf Introduction}
\smallskip
\noindent
There are at least two good reasons for testing the validity of
Einstein's General Relativity (EGR) and its weak field limit,
Newtonian Gravity (NG), over cosmic distances;

(a) It has not been tested before over such distances -
 All astronomical tests of EGR and NG, so far, were limited to
the solar system$^1$ and to close binary systems$^{2-6}$
(PSR 1913+16, 4U1820-30 and PSR 0655+64), i.e.,
to distance scales less than 50 Astronomical Units
whereas EGR and NG have been applied to astronomical systems such as
galaxies, clusters of galaxies, superclusters and the whole
Universe, which are typically $10^6-10^{15}$ times larger.

(b) The dark matter problem -
All the dynamical evidence from galaxies,
clusters of galaxies, superclusters and large scale structures
that they contain vast quantities of non luminous dark matter$^7$
has been obtained using EGR or
NG for such large astronomical systems.
Thus far, in spite of extensive laboratory searches
no conclusive evidence has been found
for finite neutrino masses or for the existence of other dark
matter particles beyond the standard model that can solve the
dark matter problem. This
has led some authors to question the validity of EGR and NG
over large distances and to suggest$^9$ that perhaps EGR and NG
are only approximate theories of gravity and that a correct
theory of gravity will eliminate the dark matter problem. Indeed
alternative theories$^{10}$ to General Relativity or
modifications$^{11-13}$ of Newton's laws have been proposed
in order to explain the observations without invoking dark matter.
\smallskip

The general success of EGR
in explaining observations of gravitational lensing of quasars
and galaxies by distant galaxies and clusters of galaxies$^{14}$
suggests that it is valid over cosmological distances.
However, most of the theoretical studies of gravitational lensing were
devoted to interpreting detailed gravitational lensing observations
and to extracting astrophysical and cosmological information from these
observations. In all these studies EGR has been assumed to be the correct
theory of gravity but was not tested explicitly. In this note, however,
I use only general features of gravitational lensing of distant
quasars and galaxies by galaxies or cluster of galaxies,
which are not sensitive to lens modelling,
to test directly the validity of EGR and
Newton's laws over galactic and intergalactic distances.
The tests are similar to the historical 1919 and 1967
tests of EGR from the deflection of light by  the Sun$^{15,16}$
and from the time delay suffered by
radar signals in the gravitational field of the Sun$^{17,18}$. Both
in these historical solar system tests of EGR and in the
galactic tests of EGR, one compares the measured and the predicted
deflection and/or time delay
of light by the gravitational field of a body whose mass has been
determined from Kepler's third law applied to the motion of a test
particle around it (the Earth around the Sun, a star, a globular
cluster or a gas cloud around the center of a galaxy).
We show that the simplest known cases of gravitational lensing of
distant sources (quasars or faint blue galaxies) by galaxies
and clusters of galaxies near the line of sight to these sources
confirm the validity of EGR and its weak field limit, NG, over
galactic and intergalactic distances and the existence of
large quantities of dark matter in galaxies and clusters of galaxies.
Alternative theories of gravity that claim to solve the dark matter
problem must pass both the precision tests of EGR
(the solar system tests  and the tests from pulsar timing in close binary
systems) and explain the observations of gravitational lensing
of quasars and galaxies by distant galaxies and clusters of galaxies .
\bigskip
{\bf II. Gravitational Lensing }
\smallskip
{\bf A. Gravitational Light Deflection}:
Einstein's theory of general relativity predicts that light which
passes at an impact parameter $b$ from a spherical symmetric mass
distribution is deflected
by an angle which, for small angles, is given approximately by
$$ \alpha\approx {4GM(b)\over c^2b}~, \eqno \eq$$
where G is Newton's gravitational constant and $M(b)$ is the
mass interior to $b$.  The mass $M(r)$ enclosed within a radial distance
$r$ from the center is given by Kepler's third law
$ M(r)\approx v_{cir}^2r/ G~, $
where $v_{cir}$ is the circular velocity of a mass orbiting at
a distance $r$ from the center. Consequently, spiral
galaxies, which have flat rotation curves $(v_{cir}\approx const.)$
have $M(r)\propto r$, $\rho(r)\propto
1/r^2~,$  and $M(b)/b \approx \pi v_{cir}^2/2G~, $ which give
rise to a constant deflection angle independent of impact parameter,
$$ \alpha=2\pi \left ({v_{cir}\over c}\right )^2~.  \eqno\eq $$
For large spiral galaxies,
$v_{cir}\sim 250~km~s^{-1}$ and $\alpha\sim 1"~.$  In elliptical
galaxies, or clusters of galaxies, whose total mass distributions are
well described by singular isothermal sphere distributions
$\rho(r)\approx (1/2\pi G)(\sigma_{_\parallel}/c)^2~,$ the squared
circular velocity is replaced by $v_{cir}^2=2\sigma_{_\parallel}^2~,$
where $\sigma_{_\parallel}$ is the one-dimensional line-of-sight
velocity dispersion in the galaxy or the cluster, respectively.
For a typical large elliptical galaxy with $\sigma_{_\parallel}\sim
200~km~s^{-1}$ the constant deflection angle is $\alpha\sim 1.5"$
while for a rich cluster with $\sigma_{_\parallel} \sim 1000~km~s^{-1}$
the constant deflection angle is $\alpha\sim 30".$ Hence, the
large optical telescopes, VLA and/or VLBI radio telescopes can be
and have been used to discover and study gravitational lensing of
quasars and galaxies by galaxies and clusters of galaxies.

EGR and Newton's laws can be tested over
galactic and intergalactic distances by comparing the deflection
of light which is extracted from these observations
and the deflection of light which is predicted from the measured
rotation curves or line-of-sight velocity dispersions in these systems,
using the geometrical relation
$$ \vec\theta_{_I}=\vec\theta_{_S}+{D_{LS}\over D_{OS}}\vec\alpha~,
\eqno\eq$$ where
$\vec\theta_{_S}~$ and $\vec\theta_{_I}$ are the angular positions
of the source and the image, respectively, relative to the lens, and
$D_{OS}$ and $D_{LS}$ are the angular diameter distances from the
observer to the source and from the lens (the galaxy or the cluster
of galaxies) to the source, respectively.

Thus, in order to extract the deflection
angle from the angular positions of the source images one must
know both the distance ratio $D_{LS}/D_{OS}$ and the angular position of
the source relative to the lens.
In a Friedmann-Robertson-Walker
universe the angular diameter distance between
object A with redshift $z_{_A}$ and object B with redshift $z_{_B}$
is given by
$$D_{A,B}={2c\over H_0}{(1-\Omega_0-G_{A}G_{B})(G_{A}-G_{B})\over
          \Omega_0^2(1+z_{_A})(1+z_{_B})^2}~, \eqno\eq $$
where $G_{A,B}\equiv(1+\Omega_0 z_{_{A,B}})^{1/2},$
$H_0\equiv 100h~km~s^{-1}Mpc^{-1}$ is the Hubble constant and
$\Omega_0\equiv\rho_0/\rho_c$ is the present density
of the universe in critical density units, $\rho_c\equiv 8\pi G/3H_0^2~.$
Although the distance for large redshifts objects depends
strongly on the cosmological model, the ratio $D_{LS}/D_{OS}$
is independent of $H_0$ and
depends rather weakly on the cosmological model
if $z_{_L}\ll z_{_S}~,$ and $\Omega_0+\lambda_0\lsim 1.$
\smallskip
The angular position of the source must
be deduced from the multiple image pattern (angular positions and
relative magnifications) of the source which is produced by the lens.
Generally, this requires a complicated inversion procedure
and additional assumptions. However, for testing EGR
we selected only the gravitational lensing cases where the
lens is simple, the pattern-recognition is straightforward
and the deflection angle can be read directly
from the simple multiple image pattern:
\smallskip
{\bf B. Einstein Rings, Crosses, and Arcs}:
On the rare occasion that a lensing galaxy with a radially symmetric
surface density happens to lie on the line-of-sight to a distant quasar
it forms in the sky a ring image (Einstein Ring$^{19,20}$) of the
quasar around the center of the lensing galaxy, whose angular diameter is
$$\Delta \theta=4\theta_r\approx 2{D_{LS}\over D_{OS}}\alpha
\approx 4\pi {D_{LS}\over D_{OS}}\left({\sigma_{_\parallel}\over c}
\right)^2~.\eqno\eq $$
Five Einstein rings,
MG1131+0456, 1830-211, MG1634+1346, 0218+357 and MG1549+3047
were discovered thus far by high resolution radio observations$^{14}$
(see for instance Fig. 2),
but, only for MG1634+1346 are the redshifts of both the lens and the
ring image known, allowing a quantitative test of EGR.
(When the source is slightly off center, the ring breaks into a pair of
arcs, as actually observed for the ring image MG1634+1346 of a radio
lobe of a distant quasar$^{21,22}$.)
\smallskip
When the lens has an elliptical surface density and the
line of sight to the source passes very near its center,
the Einstein ring degrades into an ``Einstein Cross'', i.e.,
four images that are located
symmetrically along the two principal axes$^{14}$
(and a faint fifth image at the center), with a mean angular
separation between opposite images given approximately
by Eq.5, as observed$^{23-27}$ in the case of Q2237+0305 (Fig. 3a).

\smallskip
When an extended distant source, such as a galaxy, lies on a cusp
caustic behind a giant elliptical lens, such as a rich cluster
of galaxies, it appears as an extended luminous arc
on the opposite side of the lens$^{28,29}$. The angular distance
of the arc from the center of the lens
is given approximately by the radius of the Einstein ring.
Giant arcs were discovered, thus far, in the central regions of 13
rich clusters$^{14}$ (see for instance Fig. 4)
and in six cases, Abell 370, 963 and 2390, Cl0500-24,
Cl2244-02 and Cl0024+1654 the redshifts of both the giant arc image
and the cluster are known and the velocity dispersion in the cluster
has been estimated from the redshifts of the member galaxies or the
X ray emissin, allowing a quantitative test of EGR.
\smallskip
{\bf C. Gravitational Time Delay}:
In the thin lens approximation the time delay predicted by EGR
is a sum of the time delay due to the difference in
path length between deflected and undeflected light rays
and the time delay due to the
gravitational potential felt by the light rays$^{14}$
$$ \Delta t\approx (1+z_{_L})\left[{D_{OL}D_{OS}\over 2c D_{LS}}
(\vec{\theta}_{_I}-\vec{\theta}_{_S})^2~- {\phi(\theta_{_I})
\over c^3}\right]~, \eqno\eq $$
where $\phi(\theta_{_I})$ is the gravitational potential
of the lens at $\theta_{I}$. Thus, the time delay between two images A,B,
due to a lensing galaxy with nearly spherical isothermal mass distibution
that lies near the line-of-sight to the source
(even if it is embedded
in a large cluster with an approximately constant deflection angle
over the whole image), reduces to a simple form,
$$\Delta t_{A,B}\approx 2\pi(1+z_{_L})~
(\vert\vec{\theta}_A\vert -\vert\vec{\theta}_B\vert)\left({
\sigma_{_\parallel}\over c}
\right)^2 {D_{OL}\over c}~, \eqno\eq $$
which can also be written as $$\Delta t_{A,B}\approx (1+z_{_L})
(\vert\vec{\theta}_A\vert -\vert\vec{\theta}_B\vert)
\vert\vec{\theta}_A -\vec{\theta}_B\vert {D_{OS} \over D_{LS}}
 {D_{OL}\over 4c}~. \eqno\eq $$
Expression 8 is still valid when the lensing galaxy is embedded
in a large cluster with an approximately constant deflection angle
over the whole image.
Note that while the deflection angle is dimensionless, i.e.,
depends only on dimensionless parameters, the time delay is
dimensionfull and depends on the absolute value of the Hubble
parameter (throgh $D_{OL}$).
\bigskip
{\bf III. Gravitational Lensing Tests of EGR}
\smallskip
Figure 1 summarizes our comparison between the above
EGR predictions and  observations on the most simple known
cases of gravitational lensing of quasars and galaxies
by galaxies or cluster of galaxies. The error bars are statistical
only and do not include model uncertainties.
The comparison is described below in detail:
\smallskip
{\bf A. The Einstein Ring MG1654+1346}:
The ring image MG1654+1346 of a radio lobe of a distant
quasar at redshift $z_{_S}=1.75$, formed by a bright elliptical galaxy
at redshift $z_{_L}=0.254~$, has$^{21,22}$ an angular diameter
$\Delta\theta=1.97"\pm 0.04".$
The redshifts of the ring and lensing galaxy yield
a distance ratio $D_{LS}/D_{OS}\approx 0.73\pm 0.01~.$
Unfortunately, no published measurements are available
of the one-dimensional velocity dispersion in the lensing
galaxy. However, Langston et al.$^{21,22}$ measured a total B-band
luminosity of the
galaxy, $L_B=(1.87\pm 0.18)\times 10^{10}h^{-2}L_\odot~,$ where
$h$ is the Hubble parameter in units of 100 $km~s^{-1}~Mpc^{-1}$
and $L_\odot$ is the luminosity of the Sun. The luminosities of
bright elliptical galaxies were found by Faber-Jackson and Dressler
to be correlated with the
velocity dispersion within the inner few kiloparsecs of the galaxies
$^{30,31}$ yielding for
MG1654+1346 a velocity dispersion of $\sigma_{_\parallel}
\approx 225\pm 22~ km~s^{-1}$. Thus, Eq.5 predicts an angular
diameter of $\Delta\theta\approx 2.12"\pm 0.42"$
for the Einstein ring, in good agreement with the
observed diameter of $\Delta\theta=1.97\pm 0.04".$  A good description
is also obtained for the detailed light pattern of MG1654+1346
assuming a constant mass to light ratio for the lens inside
the Einstein ring.
\smallskip
{\bf B. The Einstein Cross Q2237+0305}:
This "Einstein Cross" is formed by a quasar with a redshift
$z_{_S}=1.695$ that lies behind the Zwicky galaxy that has a
redshift $z_{_L}=0.0394$ and a line-of-sight
velocity dispersion of$^{32}$
$209\pm 19~km~s^{-1}.$ The redshifts of the lensed quasar and the lensing
galaxy yield a distance ratio $D_{LS}/D_{OS}\approx 0.95\pm 0.01~.$
The diameter of the ring is predicted by Eq.5 to be
$\Delta\theta\approx 2.3"\pm 0.5"~.$
It agrees with the high resolution observations
of Yee$^{24,25}$ which gave $\Delta \theta=1.82"\pm 0.03"~,$ and
with recent observations with the Hubble Space Telescope (Fig. 3a)
with the JPL wide field camera which gave$^{28}$
 $\Delta\theta=1.85"\pm 0.05".$
More accurate calculations of the image positions$^{28}$ which assume
a mass distribution within the lensing galaxy proportional
to its light distribution, yield their positions within $2\%$
(see Fig. 3b), if the mass to blue light ratio is $M/L_B=12.3h$,
in good agreement with the usual observed  mass to blue light ratio,
$M/L_B=(12\pm 2)h$, in E and SO galaxies$^{33}$.
\smallskip
{\bf C. The Einstein Arc In A370}: The giant arc
subtends an angle larger than $60^0$ at a radius
of $\theta_r=26"\pm2"~$ from the center of this rich cluster.
The lensing cluster A370 and the giant arc
have redshifts$^{34,35-37}$
$z_{_L}=0.374~$ and $z_{_S}=0.724~,$ respectively,
yielding a distance ratio $D_{LS}/D_{OS}=0.392\pm 0.008~.$
Soucail et al. obtained a line-of-sight velocity dispersion for A370
of$^{35}$ $\sigma_{_\parallel}=1700\pm 170~km~s^{-1}.$ Fort et al.
found$^{37}$ $\sigma_{_\parallel}= 1340^{+230}_{-150}~km~s^{-1},$
while Henry and Lavery
found$^{38}$ $\sigma_{_\parallel}= 1587^{+360}_{-214}~km~s^{-1}.$
The weighted mean value of these measurements is
$\sigma_{_\parallel}= 1542^{+140}_{-~90}~km~s^{-1},$ which is
also consistent with the value
$\sigma_{_\parallel}\approx 1500~km~s^{-1}$ that was
estimated$^{36}$ from the total X-ray luminosity$^{39}$ of A370.
Using the mean value, Eq.5 predicts an angular radius of
$\theta_r\approx =26.2"\pm 7.1" $, in good agreement with the
observed radius,  $\theta_r\approx 26"\pm 2"~.$
\smallskip
{\bf D. The Einstein Arc In Cl2244-02}:
The giant circular arc near the center of the rich cluster
Cl2244-02, discovered by Lynds and Petrosian$^{40}$
subtends an angle of more than $110\deg$ at an angular
distance of $\theta_r=9.9\pm 0.2"~$ from the center of the cluster.
The cluster and giant arc have redshifts
$^{41,42}$ $z_{_L}=0.328~$ and $z_{_S}=2.238~,$ respectively,
yielding a distance ratio $D_{LS}/D_{OS}=0.70\pm 0.02~.$
No published data are available yet on the velocity dispersion in
Cl2244-02. However, it
can be estimated from its measured X-ray luminosity,
using a phenomenological relation between the X-ray luminosity
and the velocity dispersion in rich clusters$^{43}$.
The X-Ray luminosity of Cl2244-02 was found to be$^{44}$
$L_x\approx 3\times 10^{44}h^{-2}~erg~s^{-1}$ in the [0.5 - 4.5]
keV band, yielding a velocity dispersion of
$\sigma_{_\parallel}\approx 775 \pm 80 ~km~s^{-1}.$
With this  value Eq. 5 predict an angular radius of
$\theta_r\approx 12"\pm 3.1"~,$
in good agreement with the observed radius, $\theta_r\approx 9.9"
\pm 0.3"~.$
\smallskip
{\bf E. The Einstein Arc In Cl0024+1654}:
The giant arc discovered by Koo$^{45}$
near the center of the cluster Cl0024+1654,
subtends an angle of more than $40^0$ at an angular
distance of $\theta_r\approx 35"~$ from the center of the cluster.
The cluster has a redshift of $z_{_L}=0.39$. The redshift
of the the arc $z_{_S}$ is not known, but it is estimated
to be between 1 and 2, based on the absence of any emission lines
in the spectrum$^{46}$
yielding a distance ratio $D_{LS}/D_{OS}=0.48-0.64~.$
The line-of-sight velocity dispersion in
Cl10024+1654 was estimated$^{47,48}$
 to be $1290\pm 100~ km~ s^{-1}$, yielding
a radius of $\theta_r\approx 20"-36",$
in agreement with the observed radius $\theta_r\approx 35"$ provided that
the image has a redshift close to 2.  (The segmentation of
the arc is probably due to two galaxies near its central segment
which perturb locally the smooth singular isothermal sphere potential
but do not change significantly the average
radial distance of the arc from the center of the cluster).
\smallskip
{\bf F. Time Delay In The Double Quasar Q0957+561}:
Thus far an observed time delay between different images of the same
quasar has been established only for the double quasar Q0957+561.
Although the lens system is not completely understood$^{14}$,
an approximate test of EGR is still possible.
The lens is essentially a giant elliptical galaxy embedded in a large
cluster$^{49,50}$ and produces two quasar
images, A,B, at angular distances$^{51}$ $\vert\vec\theta_A\vert
\approx 5.218"$ and $\vert\vec\theta_B\vert\approx 1.045",$
respectively,
from the center of the lensing galaxy, with an angular separation
$\vert\vec\theta_A-\vec\theta_B\vert\approx 6.175"~.$
The lensing galaxy has a redshift $z_{_L}=0.355$
and the images have a redshift $z_{_S}=1.41$ which yield a
distance ratio $D_{OL}D_{OS}/D_{LS}\approx (0.365\pm 0.015)c/H_0$.
{}From the predicted (Eq.8) and the observed$^{51-56}$
time delay
between the two images, $\Delta t=1.48\pm 0.03~Y$ and$^{57,58}$
$\Delta t=1.1-1.3 ~Y~,$
if it is due to the giant galaxy, one obtains
$$H_0\approx (76\pm 4)(1Y/\Delta t_{A,B})\approx
50-70~km~s^{-1}~Mpc^{-1} .\eqno\eq $$  This value is consistent
with the known value of $H_0$ from a variety of other measurements$^{59}$
 ($\approx  50-80~km~s^{-1}~Mpc^{-1}$).
Detailed modeling of the lens yields similar results$^{14}$.
\endpage
{\bf IV. Mass To Light Ratios}
\smallskip
The mass interior to an Einstein ring, i.e., interior to
$b=\theta_r D_{0L}$ that follows from Eqs. 1-3 is given by
$$M(b)\approx {c^2\theta_r^2 D_{OL}\over 4G}{D_{OS}\over D_{LS}}.
\eqno\eq $$  From Eq.10 it follows that
$M(b)\approx (1.12\pm 0.04)\times 10^{10}h^{-1}M_\odot$
interior to $b\approx 0.5h^{-1}kpc$ for 2237+0305, and
$M(b)\approx (9.0\pm 0.4)\times 10^{10}h^{-1}M_\odot$ interior to
$b\approx 2.5h^{-1}kpc$ for MG1654+1346.
These masses and the measured luminosities inside the rings give
$M/L_B\approx (16\pm 2)h$ for MG1654+1346 and
$M/L_B\approx (10\pm 3)h$ for Q2237+0305 (in $ M_\odot/L_\odot$ units)
 in good
agreement with the best dynamical $M/L_B$ measurements for the cores of
bright elliptical galaxies$^{33}$, $M/L_B=(11.8\pm 2)h$.

Although the observations of 2237+0305 and MG1654+1346 confirm the
validity of EGR and Newtonian gravity for small galactic distances,
and yield gravitating masses equal, within error bars, to
the dynamical masses, i.e., to the masses derived from Newtonian
dynamics,
they shed no light on the dark matter problem because of the
relatively small impact parameters that were probed. Rotation
curves and X-ray studies of individual galaxies
suggest that most of the dark matter in galaxies lies
beyond such impact parameters$^{61}.$
In order to explore
dark matter one must examine more massive and more
distant lenses.
Such cases are provided by 957+561, A370, Cl2244-02 and Cl0024+1654.
\smallskip
{\bf Q937+561}: If the splitting between the two images of Q957+561 is
due totally to the deflecting galaxy, then
$$M(b)\approx {c^2~\theta_{A}~
   \vert\vec{\theta}_A-\vec{\theta}_B\vert~
D_{OL}\over 8G}{D_{OS}\over D_{LS}}\approx (2.1-2.3)\times
10^{12}h^{-1}M_\odot
\eqno\eq $$
interior to impact parameter $b\approx
D_{OL}\vert\vec\theta_A\vert\approx  (17\pm 0.8)h^{-1}kpc.$
The V-magnitude of the
lensing galaxy yields$^{62}$ a total luminosity of $L\approx
7.6\times 10^{10}h^{-2}L_\odot,$
and a mass-to-light ratio of $M/L\approx
23h~$ interior to $b\approx 17 h^{-1}kpc.$
\smallskip
The relative time delay and the angular positions of the two images
yield a mass
$$ M(b)\approx {c^3\Delta t\over 2G(1+z_{_L})}{\theta_{A}\over
  \vert \vert\vec{\theta}_A\vert-\vert\vec{\theta}_B\vert\vert}
\approx 4.45 \times 10^{12}M_\odot~\eqno\eq $$
and a mass-to-light ratio of $M/L\approx
58h^2$ interior to impact parameter
$b\approx 17 h^{-1}kpc.$

The analysis of velocity dispersions from stellar absorption
measurements in elliptical galaxies
leads to values of $M/L\lsim 14h$
independent of radius out to one effective radius$^{61}.$
HI rotation curves have been studied up to radii of about
$16h^{-1}kpc~$ for a few hydrogen
rich ellipticals. They indicate$^{61}$ mass-to-light ratios
somewhat larger, but not significantly larger, approaching these radii.
X-ray studies of giant ellipticals
gave$^{61}$ $M/L > 40h$ for radii between
$16h^{-1}kpc$ and $ 50h^{-1}kpc.$ Thus,
the lensing mass of 957+561 within $17h^{-1}kpc$
and the mass-to-light ratio are consistent with those derived
from HI rotation curves and X-ray studies of giant elliptical
galaxies and they
provide evidence for a large amount of dark matter in
957+561, in particular if $H_0\gg 50~km~s^{-1}Mpc^{-1}.$
\smallskip
{\bf A370}:
The gravitating
mass enclosed within the impact parameter $b=D_{OL}\theta_r
\approx (85\pm 5)h^{-1} kpc~$ from the center of the cluster
that follows from Eq.10 is
$ M(b)\approx (1.54\pm 0.10)\times 10^{14}h^{-1}M_\odot ~.$
This value yields a mass-to-light ratio
of $M/L_R\approx 130h$ within
$(85\pm 5)h^{-1}kpc~.$
Taking into account the difference between the mean galactic B - R
index and the solar one, it gives $M/L_B\approx 350h.$
Similar ratios were obtained from the virial theorem for the centers
of other rich clusters indicating the existence of
large quantities of gravitating dark matter in rich clusters of galaxies.
\smallskip
{\bf Cl2244-02}: The gravitating
mass enclosed within the impact parameter $b=D_{OL}\theta_r
\approx (31\pm 2)h^{-1} kpc~$ from the center of the cluster which
follows from Eq. 10 is:
$ M(b)\approx (1.15\pm 0.10)\times 10^{13}h^{-1}M_\odot ~.$
It yields a mass-to-light
ratio of about $M/L_V\approx 200h$ in good
agreement with the mass-to-light ratios found in other similar
clusters from the virial theorem and from X-ray observations.
Like in the case of A370, this mass-to-light ratio indicates
that the cluster contains a large amount of gravitating dark matter.
\bigskip
{\bf V. Conclusions}
\smallskip
Observations of the six most simple known cases of gravitationally
lensed images of quasars and galaxies, the Einstein Cross Q2237+05,
the Einstein Ring MG1654+1346, the double quasar Q0957+561 and
the Einstein Arcs in A370, Cl2244-02, and Cl0024+1654
confirm the predictions of
Einstein's theory of General Relativity for the deflection
and time delay in the gravitational field of galaxies and clusters
of galaxies with masses equal to their dynamical masses (masses
deduced from observed velocities using Newton's laws). These observations
confirm within error bars the consistency between
Einstein's General Relativity and Newtonian dynamics within
impact parameters of $~\sim 0.5h^{-1},~2.5h^{-1},~
17h^{-1}~$, $~31h^{-1},~85h^{-1}$ and $117h^{-1}~kpc~,$ respectively.
The lensing masses in  galaxies and
clusters are the same within error bars as their dynamical masses
and are similar to the masses of other similar galaxies
and clusters. The mass-to-light ratios in the lenses
indicate the existence of large amounts of gravitating dark
matter in the lenses within impact parameters
larger than $10h^{-1}~kpc.$ In fact
the gravitational lensing tests of EGR confirm
that our Universe consists
mainly of a mysterious gravitating dark matter.
Detailed theoretical and observational studies of gravitational lenses
can both improve the precision
of the tests of EGR at large distances and shed some light
on the distribution$^{62}$ and the nature of dark matter in galaxies,
in clusters of galaxies and in the intergalactic space.

\endpage
\centerline{{\bf REFERENCES}}
\bigskip
1. See for instance Weinberg, S., "Gravitation And Cosmology"
   (John Wiley, New York 1972) and references therein.

2. Hulse, R.H. Taylor, J.H., Ap. J. Lett. {\bf 195}, L51 (1975).

3. Will, C.M., Physics Reports {\bf 113}, 345 (1984) and references
   therein.

4. Backer, D.C. and Helings, R.H., Ann. Rev. Astron. Astrophys.
   {\bf 24}, 537 (1986) and references therein.

5. Taylor, J.H. and Weisberg, J.M., Ap. J. {\bf 345}, 434 (1989).

6. Damour, T. and Taylor, J.H., Ap. J. {\bf 366}, 501 (1991).

7.  Kolb, E.W. and Turner, M.S., "The Early Universe" (Addison-
    Wesley Pub. Com. 1990) and references therein

8. Olive, K. et al., Phys. Lett. {\bf B236}, 454 (1990).

9.  Sanders, R.H., Astron. Astrophys. Rev. {\bf 2}, 1
    (1990) and references therein.

10. See for instance Mannheim, P.D. and Kazanas, D.
    Ap. J. {\bf 342}, 635 (1989).

11. Milgrom, M., Ap. J.{\bf 270}, 365(1983); {\bf 270}, 371 (1983);
    {\bf 270}, 384 (1983); {\bf 287}, 571 (1984); {\bf 302}, 617
    (1986); {\bf 333}, 684 (1988).

12. Milgrom, M. and  Bekenstein, J., Ap. J. {\bf 286}, 7 (1984).

13. Sanders,  R.H., Asstron. Astrophys. Lett. {\bf 136}, L21 (1984).

14. See for instance Blandford, R.D. and Narayan, R., Ann. Rev. Astron.
    and Astrophys. {\bf 30}, 311 (1992).

15.  Dyson, F.W., Eddington, A.S. and Davidson, C., Phil. Trans.
     Roy. Soc. (London) {\bf 220A }, 291 (1920);
     Mem. Roy. Astron. Soc., {\bf 62}, 291 (1920).
     For recent tests see ref. 30.

16.  Robertson, D.S., Carter, W.E. and  Dillinger, W.H.,
     Nature {\bf 349},768 (1991) and references therein.

17.  Shapiro, I.I. Phys. Rev. Lett.  {\bf 13}, 789 (1964);

18.  Shapiro, I.I. et al., Phys. Rev. Let. {\bf 20}, 1265 (1968).

19. Cholson, O., Astron. Nachr. {\bf 221}, 329 (1924).

20. Einstein, A., Science {\bf 84}, 506 (1936);

21. Langston, G.I. et al., Astr. J. {\bf 97}, 1283 (1989)

22. Langston, G.I.  et al., Nature {\bf 344}, 43 (1990).

23. Schneider, D.P. et al., Astron. J. {\bf 95}, 1619 (1988).

24. Yee, H.K.C., Astr. J. {\bf 95}, 1331(1988)

25. De Robertis, M.M. and  Yee, H.K.C., Ap. J. Lett.
    {\bf 332}, L49 (1988).

26. Adams, . et al., Astron. Astrophys. {\bf 208}, L15 (1989).

27. Rix, H.W., Schneider, D.P. and Bahcall, J.N., Astron. J.
    (1992).

28. Grossman, S.A. and Narayan, R., Ap. J. Lett. {\bf 324}, L37 (1988).

29. Blandford, R.D. et al., Science {\bf 245}, 824 (1989).

30. Faber, S.M. and Jackson, R.E., Ap. J. {\bf 204}, 668 (1976).

31. Dressler, A. et al., Ap. J. {\bf 313}, 42 (1987).

32. Foltz, C. et al., Ap. J. Lett. {\bf 386}, L43 (1992).

33. van der Marel, R.P., M.N.R.A.S. {\bf 253}, 710 (1991).

34. Soucail, G. et al., Astro. Astrophys. Lett. {\bf 172}, L14 (1987).

35. Soucail, G. et al., Astron. Astrophys. Lett. {\bf 172}, L14 (1988):
    {\bf 184}, L7 (1988); {\bf 191}, L19 (1988).

36. Mellier, Y. et al., Astron. Astrophys. {\bf  }, 13 (1988).

37. Fort, B. et al., Astron. Astrophys. Lett. {\bf 200}, L17 (1988).

38. Henry, J.P. and Lavery, R.J., Ap. J. {\bf 323}, 473 (1987).

39. Henry, J.P. et al., Ap. J. {\bf 262}, 1 (1982).

40. Lynds, R. and Petrosian, V., Ap. J. {\bf 336}, 1 (1989).

41. Hammer, F. et al., Astron. Astrophys. Lett. {\bf 208}, L7 (1989).

42. Miller, J.S.  and Goodrich, R.W., Nature {\bf 331}, 685 (1988).

43. Quintana, H. and Meinick, J., Astron. J. {\bf 87}, 972 (1982).

44. Henry, J.P. et al., Ap. J. {\bf 262}, 1 (1982).

45. Koo, D., in Large Scale Motions in the Universe (ed. V.C. Rubin
    and G.V. Coyne, Princeton Univ. Press) p. 513 (1988).

46. Mellier Y. et al., Ap. J. {\bf 380}, 334 (1991).

47. Desssler, A., Schneider, D.P. and Gunnn, J.E., Ap. J. {\bf 294},
    70 (1985).

48. Schneider, D.P., Dressler, A. and Gunn, J.E., Astron. J. {\bf 92},
    523 (1986).

49. Stockton, A., Ap. J. Lett. {\bf 242}, L141 (1980)

50. Young, P. et al., Ap. J. {\bf 241}, 507 (1980).

51. Falco, E.E., Gorenstein, M.V. and Shapiro, I.I.,
    Ap. J. {\bf 372}, 364 (1991).

52. Florentin-Nielsen, R., Astron. Astrophys. {\bf 138}, L19 (1984).

53. Lehar. J. et al. Ap. J. {\bf 384}, 453 (1992).

54. Press, W.H. et al. Ap. J. {\bf 385}, 404 (1992).

55. Press, W.H., et al. Ap. J. {\bf 385}, 416 (1992).

56. Beskin, G.M. and Oknyanski, V.L.  1992

57. Vanderriest, C.  et al., Astron. Astrophys. {\bf 215} 1 (1989).

58. Schild, R.E., Astron. J. {\bf 100}, 1771 (1990).

59. van den Bergh, S., Science {\bf 258}, 421 (1992) and references
    therein.

60. Sancisi, R. and van Albada, T.S., "Observational Cosmology"
    (Reidel, Dordrecht 1987) p. 699.

61. Musshotzky, R., A.I.P. Conf. Proc. {\bf 222}, 394 (1991).

62. See for instance Tyson, J.A., A.I.P. Conf. Proc. {\bf 222}, 437
    (1991).

\endpage
\centerline{{\bf FIGURE CAPTIONS}}
\smallskip
{\bf Fig. 1: The ratios between the EGR prediction and the observations
of the deflection and time delay of light from distant quasars and
galaxies by galaxies or clusters of galaxies, displayed
at the impact parameter of the deflected light relative to the center
of the lens, for the Einstein Cross Q2237+05,
the Einstein Ring MG1654+1346, the double quasar Q0957+561 and
the Einstein Arcs in A370, Cl2244-02, and Cl0024+1654.
The estimated errors in the ratios include the quoted observational
errors and the errors in the theoretical estimates due only to
errors in measured parameters and
the absence of precise knowledge of $\Omega$ and $h~$, but not
systematic errors.
\smallskip
Fig. 2: Radio images at 15 GHz and 5 GHz of
MG1131+O456, the first discovered Einstein ring (by Hewitt et al 1988).
\smallskip
Fig. 3a: A 700 s exposure of the Einsten cross Q2237+0305
taken with the Wide Field Camera through the F336W filter on
board the Hubble Space Telescope$^{27}$.
\smallskip
Fig. 3b: A comparison between the observed (crossed boxes)
and predicted (circles) positions of
the four images of Q2237+0305 assuming a constant mass to light
ratio near the center of the lensing galaxy$^{27}$.
\smallskip
Fig. 4: The giant arc image of a distant blue galaxy formed by
the rich cluster Abell 370 is the  first discovered Einstein arc.
It was discovered independently by Lynds and Petrosian (1986) and by
Soucail, Fort, Mellllier and Picat (1987).}

\end